\documentclass[useAMS,usenatbib]{mn2e}

\usepackage{graphicx}
\usepackage{epsfig}
\usepackage{amssymb}

\def\vhel{\ifmmode{V_{{\rm HEL}}}\else{$V_{{\rm HEL}}$}\fi}
\def\vsys{\ifmmode{V_{\rm sys}}\else{$V_{\rm sys}$}\fi}
\def\kms{\ifmmode{~{\rm km\,s}^{-1}}\else{~km~s$^{-1}$}\fi}
\def\vlsr{\ifmmode{v_{\rm lsr}}\else{$v_{\rm lsr}$}\fi}

\def\ltsim{\ifmmode\stackrel{<}{_{\sim}}\else$\stackrel{<}{_{\sim}}$\fi}
\def\gtsim{\ifmmode\stackrel{>}{_{\sim}}\else$\stackrel{>}{_{\sim}}$\fi}

\def\reff@jnl#1{{\rm#1\/}}

\def\aj{\reff@jnl{AJ}}                  
\def\araa{\reff@jnl{ARA\&A}}            
\def\apj{\reff@jnl{ApJ}}                
\def\apjl{\reff@jnl{ApJ}}               
\def\aplett{\reff@jnl{ApJ}}             
\def\apjs{\reff@jnl{ApJS}}              
\def\ao{\reff@jnl{Appl.Optics}}         
\def\apss{\reff@jnl{Ap\&SS}}            
\def\aap{\reff@jnl{A\&A}}               
\def\aapr{\reff@jnl{A\&A~Rev.}}         
\def\aaps{\reff@jnl{A\&AS}}             
\def\azh{\reff@jnl{AZh}}                        
\def\baas{\reff@jnl{BAAS}}              
\def\jrasc{\reff@jnl{JRASC}}            
\def\memras{\reff@jnl{MmRAS}}           
\def\mnras{\reff@jnl{MNRAS}}            
\def\pra{\reff@jnl{Phys.Rev.A}}         
\def\prb{\reff@jnl{Phys.Rev.B}}         
\def\prc{\reff@jnl{Phys.Rev.C}}         
\def\prd{\reff@jnl{Phys.Rev.D}}         
\def\prl{\reff@jnl{Phys.Rev.Lett}}      
\def\pasp{\reff@jnl{PASP}}              
\def\pasj{\reff@jnl{PASJ}}              
\def\qjras{\reff@jnl{QJRAS}}            
\def\skytel{\reff@jnl{S\&T}}            
\def\solphys{\reff@jnl{Solar~Phys.}}    
\def\sovast{\reff@jnl{Soviet~Ast.}}     
 \def\ssr{\reff@jnl{Space~Sci.Rev.}}     
\def\zap{\reff@jnl{ZAp}}                        
\def\nat{\reff@jnl{Nature}}             

\def\LaTeX{L\kern-.36em\raise.3ex\hbox{a}\kern-.15em
    T\kern-.1667em\lower.7ex\hbox{E}\kern-.125emX}

\def\h2{{\rm H}{\sc ii}}

\begin{document}

\title[Constraints on the Polarization of Anomalous Microwave Emission in molecular clouds]{New constraints on the Polarization of Anomalous Microwave Emission in nearby molecular clouds}

\author[C. Dickinson et al.]{C.~Dickinson,$^{1}$\thanks{E-mail: Clive.Dickinson@manchester.ac.uk} M.~Peel$^{1}$ and M.~Vidal$^{1}$ \\
$^1$Jodrell Bank Centre for Astrophysics, Alan Turing Building, School of Physics \& Astronomy, University of Manchester, Oxford Rd, \\
Manchester, M13 9PL
}

\date{Received **insert**; Accepted **insert**}
       
\pagerange{\pageref{firstpage}--\pageref{lastpage}} 
\pubyear{}

\maketitle
\label{firstpage}


\begin{abstract}
Anomalous Microwave Emission (AME) has been previously studied in two well-known molecular clouds and is thought to be due to electric dipole radiation from small spinning dust grains. It is important to measure the polarization properties of this radiation both for component separation in future cosmic microwave background experiments and also to constrain dust models. We have searched for linearly polarized radio emission associated with the $\rho$~Ophiuchi and Perseus molecular clouds using {\it WMAP} 7-year data. We found no significant polarization within an aperture of $2^{\circ}$ diameter. The upper limits on the fractional polarization of spinning dust in the $\rho$~Ophiuchi cloud are $1.7\,\%$, $1.6\,\%$ and $2.6\,\%$ (at 95\,\% confidence level) at K-, Ka- and Q-bands, respectively. In the Perseus cloud we derived upper limits of $1.4\,\%$, $1.9\,\%$ and $4.7\,\%$, at K-, Ka- and Q-bands, respectively; these are similar to those found by L\'{o}pez-Caraballo et al. If AME at high Galactic latitudes has a similarly low level of polarization, this will simplify component separation for CMB polarization measurements. We can also rule out single domain magnetic dipole radiation as the dominant emission mechanism for the 20--40\,GHz. The polarization levels are consistent with spinning dust models.
\end{abstract}



\setcounter{figure}{0}

\section{INTRODUCTION}
\label{sec:introduction}
\begin{figure*}
\begin{center}
\includegraphics[width=0.9\textwidth,angle=0]{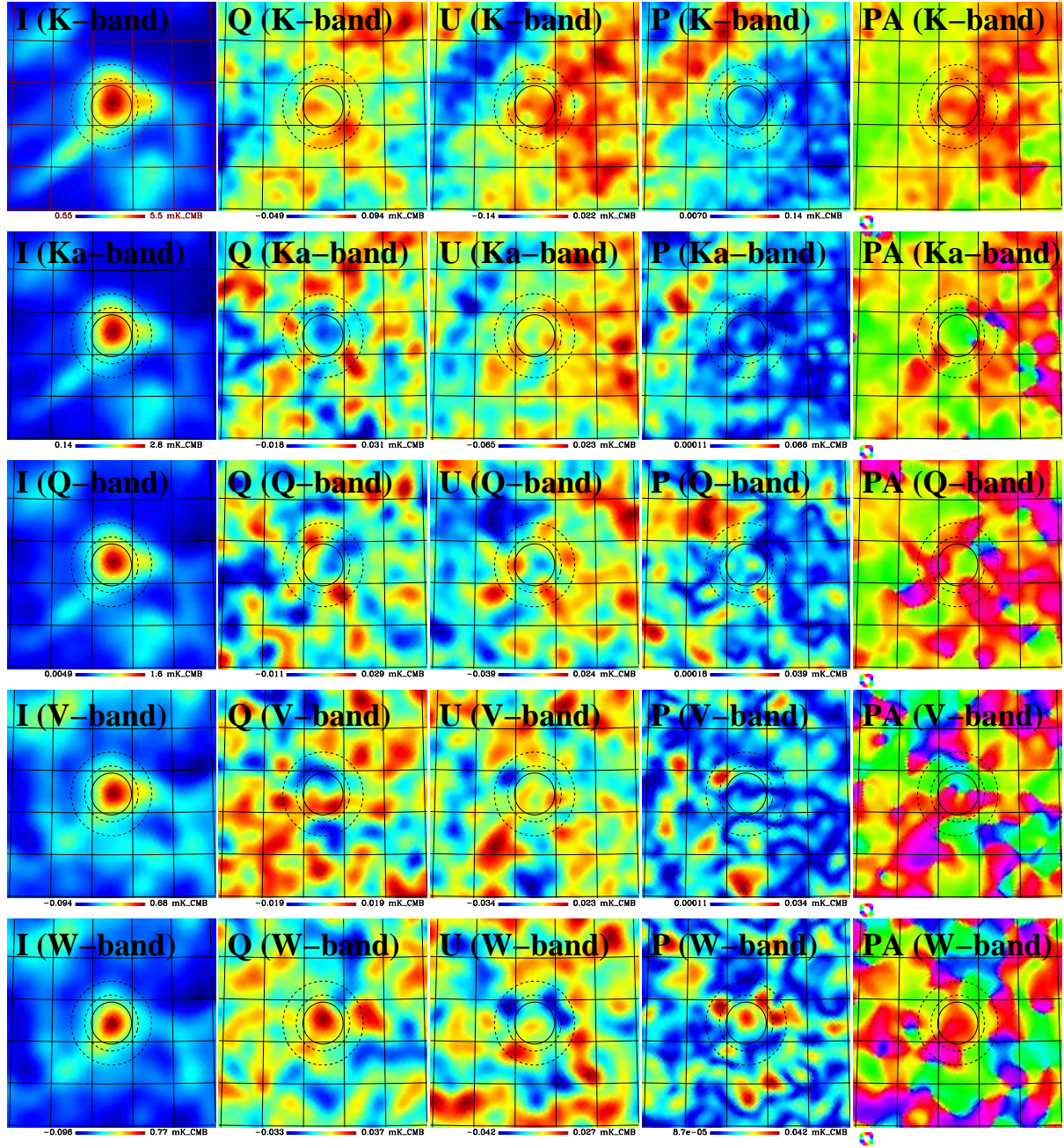}
 \caption{{\it WMAP} 7-year maps of the $\rho$~Ophiuchi region at 22.7 (K-band), 33.0 (Ka-band), 40.7 (Q-band), 60.6 (V-band), 93.4\,GHz (W-band). Each map covers $10^{\circ} \times 10^{\circ}$ centred at $(l,b)=(353.\!^{\circ}05, +16.\!^{\circ}90)$ and is smoothed to $1^{\circ}$ resolution. From left to right are Stokes $I$ (total-intensity), $Q$, $U$, polarized intensity ($P$) and polarization angle (PA). Units are thermodynamic (CMB) mK. The graticule has a spacing of $2^{\circ}$. The primary extraction aperture is shown as a solid line and the background annulus as a dashed line. \label{fig:maps}}
\end{center}
\end{figure*}
Anomalous Microwave Emission (AME) has been observed by numerous experiments in the frequency range $\sim\,10$--$100$\,GHz as an excess compared to synchrotron, free-free, cosmic microwave background (CMB) and thermal dust emissions \citep{Leitch97,Kogut96}. The AME is observed as diffuse emission at high latitudes \citep{deOliveira-Costa04,Bonaldi07,Miville-Deschenes08,Dobler08,Ysard10,Gold11} as well as in specific Galactic objects \citep{Finkbeiner02,Watson05,Casassus06,Casassus08,Dickinson09a,Dickinson10,Scaife09,Tibbs10,Castellanos11,Vidal11}. The physical mechanism responsible for the AME is a matter of debate. However, there is significant evidence for electric dipole radiation from small spinning dust grains \citep{Draine98a,Draine98b}. In particular, the recent results from {\it Planck} \citep{Planck2011-7_2} have shown striking evidence for a spinning dust spectrum for both the Perseus and $\rho$~Ophiuchi molecular clouds.

The exact amplitude of the AME signal is still not well understood since the results depend critically on the component separation of multiple diffuse components (free-free, synchrotron, thermal dust and CMB). For example, in the case of template fitting, where AME is correlated with far-infrared (FIR) maps, the AME is the strongest foreground in total-intensity in the range $\sim\,20$--$40$\,GHz \citep{Banday03,Davies06,Dickinson09b}. It is therefore crucial to understand this component in terms of CMB foreground removal. Given that the diffuse polarized foregrounds are much brighter than the CMB polarization signal \citep{Page07,Betoule09}, it is critical to understand the nature of the AME and determine its polarization properties. Although we expect synchrotron radiation to be the dominant polarized foreground at frequencies $\lesssim 100$\,GHz, polarized AME could be significant.

To date, there have been very few measurements of the polarization of AME. The first measurements were made with the COSMOSOMAS experiment at 11\,GHz in the Perseus Molecular cloud. \cite{Battistelli06} measured the fractional polarization to be $3.4^{+1.5}_{-1.9}\,\%$ at 95\,\% confidence level (c.l.). \cite{Kogut07} used the full-sky {\it WMAP} 3-year data to constrain the polarization fraction of the AME as traced by FIR maps and concluded that it contributes less than 1\,\% of the observed polarization signal variance. \cite{Dickinson06} gave an upper limit of 10\,\% ($2\sigma$) to the possible excess emission from the HII region LPH96 at 31\,GHz while \cite{Casassus07} gave an upper limit of $12\,\%$ for the excess 31\,GHz emission from the Helix nebula. \cite{Mason09} placed an upper limit of $2.7\,\%$ ($95\,\%$ c.l.) at 9.65\,GHz for LDN1622 at an angular scale of $1.3$~arcmin. \cite{Casassus08} gave a $3\,\sigma$ upper limit of $4.8\,\%$ at the peak intensity of the $\rho$~Ophiuchi CBI map at 31\,GHz, on scales of $\approx 9$~arcmin. More recently, \cite{Lopez-Caraballo11} used the {\it WMAP} 7-year data to place upper limits of 1.0, 1.8 and 2.7\,\%, at 23, 33 and 41\,GHz, respectively (with a 95\,\% c.l.), within the Perseus Molecular Cloud. \textbf{\cite{Macellari11} used a template fitting technique to constrain the dust-correlated component at 23\,GHz to be less than 5\,\%.} These measurements therefore indicate that the AME has a relatively low ($\lesssim 3\,\%$) level of polarization, which is consistent with expectations for spinning dust models based on resonance relaxation proposed by \cite{Lazarian00}. At the same time, they largely rule out models of magnetic dipole radiation \citep{Draine99}, which typically predict high polarization fractions (up to $\sim 40\,\%$). However, it should be noted that \cite{Draine99} also show a model with random inclusions of metallic Fe that produces very little polarization ($<1\,\%$).  

In this Letter, we examine the {\it WMAP} 7-year data to constrain the polarization fraction for the AME in the $\rho$~Ophiuchi and Perseus clouds. Both regions were studied in intensity using data from the {\it Planck} satellite and were shown to be dominated by AME at frequencies of $\sim 20$--$60$\,GHz \citep{Planck2011-7_2}, with spectra that can be well-fitted by models of spinning dust grains. Section~\ref{sec:maps} describes the {\it WMAP} maps used in the analysis. Section~\ref{sec:analysis} presents the analysis and a discussion of the results. Section~\ref{sec:conclusions} concludes.


\section{MAPS}
\label{sec:maps}

The analysis is based on the {\it WMAP} 7-year data available from the LAMBDA website\footnote{http://lambda.gsfc.nasa.gov/}. We used the $1^{\circ}$-smoothed maps available in HEALPix format with $N_{\rm side}=512$ \citep{Gorski05}. The maps contain the thermodynamic brightness temperature (mK$_{\rm CMB}$) with respect to the CMB temperature, at each pixel in Stokes $I$, $Q$ and $U$. The data also contain estimates of the white noise level in each pixel, including the noise correlation between $Q$ and $U$. 

Fig.~\ref{fig:maps} shows $10{^\circ} \times 10^{\circ}$ {\it WMAP} maps of the $\rho$~Ophiuchi region centred at Galactic coordinates $(l,b)=(353^{\circ}\!.05,+16.\!^{\circ}90)$ at each of the five {\it WMAP} bands (see caption of Fig.~\ref{fig:maps}). The maps depict Stokes $I$ (total intensity) and linear polarization in the form of Stokes $Q$ and $U$ maps. We also show the polarized intensity ($P$) and polarization angle (PA) maps derived directly from the Q and U maps, where $P=\sqrt{Q^2 + U^2}$ and ${\rm PA}=\frac{1}{2}\tan^{-1}(U/Q)$. In total-intensity, there is a bright feature seen in the centre, which corresponds to the $\rho$~Ophiuchi molecular cloud, or AME-G353.05+16.90 of \cite{Planck2011-7_2}. It is detected with high signal-to-noise ratio at all frequencies, with the majority of the flux falling within the $2^{\circ}$ aperture. There is fainter extended emission around the main feature and also in the surrounding region. 

\begin{figure}
\begin{center}
\includegraphics[width=0.4\textwidth,angle=0]{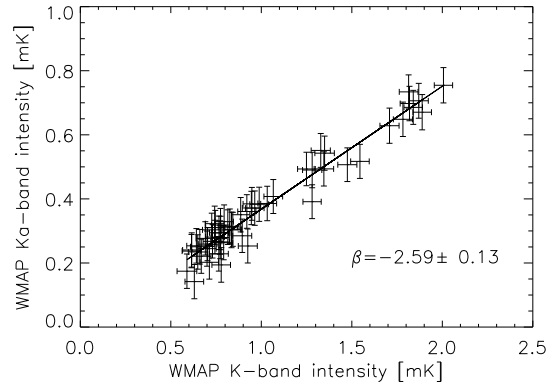}
 \caption{T-T plot of Ka-band (33\,GHz) vs K-band (22.7\,GHz) for a $5^{\circ} \times 5^{\circ}$ region centred at $(l,b)=(347.\!^{\circ}8, +16.\!^{\circ}8)$. The least-square straight line is shown. The slope corresponds to a temperature spectral index $\beta=-2.59\pm0.13$. \label{fig:TTplot}}
\end{center}
\end{figure}

In polarization, the maps look very different. There are large-scale features running across the maps, particularly at the lower frequencies of K- and Ka-bands. There is a relatively bright synchrotron spur running diagonally across the top left-hand side of the polarization maps, which is not obvious from the total-intensity maps. This is presumably because there is strong AME and free-free emission which is dominating over the synchrotron component. However, in polarization, the synchrotron component is strong while the AME/free-free components is clearly much weaker. Away from the bright spur, there is weak polarized emission in the background of $\rho$~Ophiuchi region and extending to the right-hand side of the map. This produces coherency in the polarization angle at K- and Ka-bands (see Fig.~\ref{fig:maps}). A T-T analysis (Fig.~\ref{fig:TTplot}) of the a $5^{\circ} \times 5^{\circ}$ region centred at $(l,b)=(347.\!^{\circ}8, +16.\!^{\circ}8)$ finds a spectral index between 22.7 and 33.0\,GHz of $\beta \approx -2.6$ ($T \propto \nu^{\beta}$). This is consistent with non-thermal synchrotron emission. At Q-band (40.6\,GHz) the polarized emission is weaker and the maps are dominated by noise. 

In addition to the diffuse synchrotron emission, compact (polarized) emission  may be present in the form of extragalactic radio sources; however, a search of the NASA Extragalactic Database\footnote{http://ned.ipac.caltech.edu/} found no bright ($\gtrsim 100$\,mJy at {\it WMAP} frequencies or below) radio sources within a $2^{\circ}$ radius that would be detected in the {\it WMAP} data.

\section{ANALYSIS AND DISCUSSION}
\label{sec:analysis}
\begin{table*}
\begin{center}
\caption{Aperture photometry of the {\it WMAP} 7-year maps in the $\rho$~Ophiuchi ({\it top}) and Perseus ({\it bottom}) molecular cloud regions, using a $2^{\circ}$ diameter aperture. The columns contain flux densities for intensity $I$, the spinning dust contribution to the intensity ($I_{\rm sd}$), $Q$, $U$, the observed polarized intensity ($P$), the maximum likelihood value for the noise-bias corrected polarized intensity ($P_{0}$), the total polarization fraction ($\Pi$), and the polarization fraction for spinning dust ($\Pi_{\rm sd}$). Upper limits are at the $95$ c.l.}
\begin{tabular}{lccccccccc}
\hline
Frequency          &$I$            &$I_{\rm sd}$   &$Q$              &$U$              &$P$               &$P_{0}$              &$\Pi$            &$\Pi_{\rm sd}$ \\
(GHz)              &(Jy)         &(Jy)          &(Jy)           &(Jy)           &(Jy)            &(Jy)                        &(\%)                &(\%)       \\      \hline 
22.7 (K-band)      &$26.3\pm5.5$ &$24.8\pm 6.6$ &$0.091\pm0.096$&$0.23\pm0.14$  &$0.25 \pm 0.13$ &$0.21^{+0.11}_{-0.15} (<0.43)$ &$<1.6$           &$<1.7$ \\
33.0 (Ka-band)     &$30.7\pm5.3$ &$27.2\pm 6.3$ &$-0.27\pm0.12$ &$0.02\pm0.15$  &$0.27 \pm 0.12$ &$0.24^{+0.12}_{-0.14} (<0.44)$ &$<1.4$           &$<1.6$ \\
40.7 (Q-band)     &$27.7\pm4.6$ &$21.9\pm 5.8$ &$-0.07\pm0.18$ &$-0.23\pm0.24$ &$0.24 \pm 0.23$ &$0.00^{+0.30} (<0.57)$         &$<2.1$           &$<2.6$ \\
60.6 (V-band)     &$26.3\pm4.5$ &$9.8\pm 6.5$ &$0.35\pm0.41$  &$0.41\pm0.40$  &$0.54 \pm 0.41$ &$0.00^{+0.61} (<1.1)$          &$<4.2$           &$<11$ \\
93.4 (W-band)     &$63.6\pm8.9$ &$6\pm 15$ &$2.8\pm1.0$    &$0.0\pm1.7$    &$2.8 \pm 1.0$   &$2.6^{+1.0}_{-1.1}$             &$4.1\pm1.8$      &$...$ \\
\hline
22.7 (K-band)      &$21.0\pm3.1$ &$16.7\pm3.5$ &$-0.118\pm0.071$&$0.068\pm0.069$ &$0.136 \pm 0.070$ &$0.111^{+0.060}_{-0.084} (<0.24)$ &$<1.1$           &$<1.4$ \\
33.0 (Ka-band)     &$20.4\pm3.0$ &$15.7\pm3.3$ &$0.02\pm0.13$   &$-0.06\pm0.15$  &$0.07 \pm 0.15$ &$0.00^{+0.15} (<0.30)$        &$<1.5$           &$<1.9$ \\
40.7 (Q-band)     &$16.9\pm2.9$ &$11.6\pm3.3$  &$0.07\pm0.19$   &$-0.25\pm0.21$  &$0.26 \pm 0.21$ &$0.00^{+0.30} (<0.54)$        &$<3.2$           &$<4.7$ \\
60.6 (V-band)     &$14.9\pm4.0$ &$5.4\pm4.4$  &$0.098\pm0.077$   &$-0.64\pm0.59$  &$0.65 \pm 0.39$ &$0.44^{+0.27}_{-0.44} (<1.2)$          &$<8.1$           &$<22$ \\
93.4 (W-band)     &$32.4\pm9.8$ &$3\pm10$  &$-0.5\pm1.3$    &$-0.3\pm2.3$    &$0.6 \pm 1.4$   &$0.0^{+1.5} (<2.9)$                 &$<9.0$   &$...$ \\
\hline
\end{tabular}
\end{center}
\label{tab:aperture_photometry}
\end{table*} 
We carried out aperture photometry of the maps using the same $2^{\circ}$ diameter aperture as in \cite{Planck2011-7_2}, indicated in Fig.~\ref{fig:maps}. The results are given in Table~\ref{tab:aperture_photometry}. The results for the Perseus cloud, with the same aperture centred at $(l,b)=(160.\!^{\circ}26,-18.\!^{\circ}62$), are also given. The temperatures were converted to flux density units (Jy) for the effective frequencies of {\it WMAP}. An estimate of the background was subtracted based on the median value of an annulus with an inner radius of 80~arcmin and outer radius of 120~arcmin (Fig.~\ref{fig:maps}). We also list the total-intensity flux density of the spinning dust, I$_{\rm sd}$, from \cite{Planck2011-7_2}, which appears to dominate the K-, Ka-, and Q-bands. For Perseus, we could not use the flux densities directly from \cite{Planck2011-7_2} because they filtered the data; we re-calculated the flux densities for a $2^{\circ}$ aperture\footnote{The flux densities for Perseus are lower than those of \cite{Planck2011-7_2} since they used a Gaussian model with a larger area than we used here.}. Colour corrections were applied to the flux densities for the intensity, using a model consisting of free-free, spinning dust, a CMB fluctuation, and thermal dust with fixed emissivity index and dust temperature derived by \cite{Planck2011-7_2}. These corrections were small (typically $\approx 1\,\%$ or less). No colour corrections were applied in polarization.

The uncertainty in intensity ($I$) was estimated using the r.m.s. fluctuations within the background annulus, which contains a contribution from the instrumental noise and sky background fluctuations. The code was validated by calculating the flux densities of well-known bright radio sources and comparing them to the literature. For $Q$ and $U$ we estimated the noise-only contribution using monte carlo simulations. We generated 500 noise-only simulations, based on the $Q$-$U$ noise matrices provided by the WMAP team, smoothing each noise realisation by the appropriate beam transfer function to produce $1^{\circ}$-smoothed (Gaussian FWHM) maps. In this way, we also take into account the correlated noise between the $Q$ and $U$ channels. In the $\rho$~Ophiuchi region, this corresponds to an increase of $\approx 30\,\%$ in the effective noise level. We confirmed that estimates of the background fluctuations were similar to those found in the noise-only simulations i.e. that we are noise-dominated in polarization. We also verified empirically that the background fluctuations were sub-dominant by repeating the analysis in ten random apertures in the vicinity of the $\rho$~Ophiuchi cloud; the scatter in the derived flux densities were consistent ($<3\,\sigma$) with the quoted uncertainties. 

Since we are measuring the polarization properties of Galactic signals that are relatively bright in intensity, we must be sure that instrumental polarization leakage (from $I$ to $Q$ and $U$), due to bandpass mismatch, is negligible. The {\it WMAP} $I,Q,U$ maps have been corrected for bandpass mismatch by solving for an extra term, $S$ (the ``spurious'' map), which does not vary with parallactic angle (since bandpass mismatch only depends on the intensity and spectral shape of the signal). To estimate the level of residual leakage we measured the polarization of bright Galactic H{\sc ii} regions, using the same technique described above. For M42 (Orion nebula), we constrain the polarization leakage to be $\lesssim 0.1\,\%$ at K-band and even less in the other channels, and hence this effect can be safely neglected. We therefore do not include any additional contribution to the uncertainties from systematic errors in polarization.

The observed polarized intensity was calculated as $P=\sqrt{Q^2 + U^2}$ with an uncertainty propagated from the $Q$ and $U$ uncertainties. However, it is well-known that $P$ is biased positive due to the polarization noise bias, which is particularly important at low signal-to-noise ratios \citep{Wardle74,Simmons85,Vaillancourt06}. Since it is clear that we are in the low signal-to-noise regime, we calculated confidence intervals in $P$ using a Bayesian technique, as outlined by \cite{Vaillancourt06}, and generalized to allow for non-equal noise values in $Q$ and $U$. The noise estimates from the monte carlo simulations were used to calculate the maximum likelihood value for the polarized intensity ($P_0$), along with $68\,\%$ and $95\,\%$ confidence interval, correctly taking into account the effect of noise bias. The values for both the $\rho$~Ophiuchi and Perseus molecular cloud regions are given in Table~\ref{tab:aperture_photometry}. The uncertainties are the $68\,\%$ confidence limits around the maximum likelihood values while upper limits are given at the $95\,\%$ c.l.

From the derived polarization intensity values ($P_0$), we can derive the polarization fractions for the total intensity ($\Pi=I/P_0$) and for the spinning dust component ($\Pi_{\rm sd}=I_{\rm sd}/P_0$); these are listed in Table~\ref{tab:aperture_photometry}. It can be seen that for most channels, the polarization  is an upper limit at the $95\,\%$ c.l. For $\rho$~Ophiuchi, the upper limits are at the $\lesssim 2\,\%$ level at $20$--$40$\,GHz where spinning dust is the dominant component in intensity; the tightest constraint is at 33\,GHz with an upper limit of $1.6\,\%$. At W-band ($93.4$\,GHz) there appears to be a hint of a polarization signal at $\approx 2.3\,\sigma$, which is visible in the $Q$ map (Fig.~\ref{fig:maps}). It is not clear whether this is a real detection, since W-band is more susceptible to $1/f$ noise, and therefore the noise bias/confidence interval may not be exactly correct. Nevertheless, the Rayleigh-Jeans tail of thermal dust is dominant at this frequency \citep{Planck2011-7_2} and we may be detecting the polarization of thermal dust emission, where a $\approx 4\,\%$ polarization fraction is reasonable. For the Perseus cloud, our upper limits of 1.4\,\%, 1.9\,\% and 4.7\,\% at K-, Ka-, and Q-bands, respectively, are similar to the values (1.0\,\%, 1.8\,\% and 2.7\,\%) derived by \cite{Lopez-Caraballo11}.

Such a low level of polarization of spinning dust emission is important for two main reasons. Firstly, if this can be shown to be true for diffuse emission at high Galactic latitudes, it will simplify component separation for future ultra-sensitive CMB polarization experiments. The dominant polarization foreground at frequencies $<100$\,GHz will be synchrotron emission, which is polarized at the $>20\,\%$ level at high latitudes \citep{Kogut07}. Secondly, the polarization fraction can rule out some models of dust emission. Spinning dust emission is expected to be largely unpolarized, with predictions of $<1\,\%$ at frequencies $>20$\,GHz \citep{Lazarian00}. Magnetic dipole radiation can exhibit high ($\gtrsim 10\,\%$) polarization at $20$--$40$\,GHz when there is a single magnetic domain \citep{Draine99}. This is clearly incompatible with the constraints derived here; see \cite{Lopez-Caraballo11} for a discussion. However, we note that magnetic dipole radiation may also be relatively unpolarized ($<1\,\%$) if the magnetic inclusions are randomized on the grain surface. \textbf{Also, the constraints on the polarization of AME obtained for molecular clouds may not be the same for other interstellar phases. For instance, hypothetically, one may have single magnetic dipole aligned grains in warm and hot interstellar phases, but they coagulate in molecular clouds to produce large grains with random inclusions.}


\section{CONCLUSIONS}
\label{sec:conclusions}

We have searched for polarized radio emission associated with the $\rho$~Ophiuchi molecular cloud using {\it WMAP} data. The bulk of the emission at 20--40\,GHz is due to AME, most likely from spinning dust grains \citep{Planck2011-7_2}. Thus measurements of the polarization correspond to a constraint on the polarization of spinning dust emission. We find no significant polarization within an aperture of $2^{\circ}$ diameter. The upper limits on the fractional polarization of spinning dust in the $\rho$~Ophiuchi cloud are $1.7\,\%$, $1.6\,\%$ and $2.6\,\%$ (at 95\,\% c.l.) at K-, Ka- and Q-bands, respectively. Such low levels of polarization rule out single domain magnetic dipole radiation as the dominant emission mechanism for the 20--40\,GHz signal but is consistent with spinning dust emission. Our results for the Perseus cloud are at a similar level and are comparable to those derived by \cite{Lopez-Caraballo11}. If AME at high Galactic latitudes also exhibits low ($\lesssim 1\,\%$) levels of polarization then the dominant diffuse foreground at frequencies $<100$\,GHz will be synchrotron, thus simplifying component separation for CMB polarization studies.


\section*{ACKNOWLEDGMENTS}

We thank J.\,P.\,Leahy for useful discussions about polarization noise bias and systematics in the {\it WMAP} data. CD acknowledges an STFC Advanced Fellowship and an ERC IRG grant under the FP7. MV acknowledges the funding from Becas Chile. We acknowledge the use of the HEALPix software and the use of the Legacy Archive for Microwave Background Data Analysis (LAMBDA). Support for LAMBDA is provided by the NASA Office of Space Science. This research has made use of the NASA/IPAC Extragalactic Database (NED) which is operated by the Jet Propulsion Laboratory, California Institute of Technology, under contract with the National Aeronautics and Space Administration.


\bibliographystyle{mn2e}


\bsp 

\label{lastpage}

\end{document}